\documentclass[conference]{IEEEtran}
\IEEEoverridecommandlockouts
\usepackage{cite}
\usepackage{amsmath,amssymb,amsfonts}

\usepackage{graphicx}
\usepackage{textcomp}
\usepackage{xcolor}
\def\BibTeX{{\rm B\kern-.05em{\sc i\kern-.025em b}\kern-.08em
    T\kern-.1667em\lower.7ex\hbox{E}\kern-.125emX}}

\usepackage[T1]{fontenc}
\usepackage[utf8]{inputenc}
\usepackage{amsmath,amssymb,amsthm}
\usepackage{graphicx}
\usepackage{xcolor}
\usepackage{tikz}
\usepackage{pgfplots}
\pgfplotsset{compat=1.18}
\usetikzlibrary{
  arrows.meta, positioning, shapes.geometric,
  decorations.pathreplacing, backgrounds, fit,
  calc, shadows, matrix, chains
}
\usepackage{algorithm}
\usepackage{algpseudocode}
\algnewcommand\algorithmicinput{\textbf{Input:}}
\algnewcommand\algorithmicoutput{\textbf{Output:}}
\algnewcommand\Input{\item[\algorithmicinput]}
\algnewcommand\Output{\item[\algorithmicoutput]}
\usepackage{listings}
\usepackage{booktabs, multirow, array}

\definecolor{l0col}{HTML}{546E7A}    
\definecolor{l1col}{HTML}{1565C0}    
\definecolor{l2col}{HTML}{212121}    
\definecolor{s1g}{HTML}{1B5E20}      
\definecolor{s8b}{HTML}{B71C1C}      
\definecolor{c2r}{HTML}{C62828}      
\definecolor{orb}{HTML}{37474F}      
\definecolor{pkblue}{HTML}{0D47A1}

\newtheoremstyle{defstyle}
  {6pt}{6pt}{\normalfont}{0pt}{\bfseries}{}{ }
  {\thmname{#1}~\thmnumber{#2}\thmnote{: #3.}}
\theoremstyle{defstyle}
\newtheorem{definition}{Definition}

\newtheorem{proposition}{Proposition}

\usepackage{balance}

\begin{document}

\title{Fifty Shades of Darknet\\


}
\author{
\IEEEauthorblockN{
Siddique Abubakr Muntaka\IEEEauthorrefmark{1},
Faiza Hussein\IEEEauthorrefmark{2},
Jacques Bou Abdo\IEEEauthorrefmark{1}
}
\IEEEauthorblockA{
\IEEEauthorrefmark{1}School of Information Technology, University of Cincinnati, Cincinnati, OH 45221, USA\\
\IEEEauthorrefmark{2}Garden City University College, Kumasi, Ghana\\
\{muntaksr@mail.uc.edu, bouabdjs@ucmail.uc.edu\}
}
}

\maketitle

\begin{abstract}
The Invisible Internet Project (I2P) is a peer-to-peer anonymous overlay network whose architecture includes a structurally distinct sublayer whose security implications remain uncharacterized. We term this sublayer the Exclusive Network: nodes here host operational services and draw on I2P's routing resources, but publish no RouterInfo record to the network's distributed database (NetDB). In a controlled three-node testbed, we demonstrate that an Exclusive Network node survives sequential floodfill queries from a pool of routers with zero NetDB hits, while its hosted service remains continuously accessible to authorized peers. This property is exploitable by documented I2P-based malware, for example, I2PRAT (RATatouille), for persistent command-and-control operations against national assets or corporate networks. The structure is analogous to nation-state Operational Relay Box (ORB) infrastructure. The existence of this sublayer, together with the inability of top-down empirical mapping to characterize it, motivates a move toward formal analytical methods to understand the emergence and behavior of covert networks within I2P.
\end{abstract}

\section{Introduction}
\label{sec:intro}

The Invisible Internet Project (I2P), among other anonymity networks, carries significant
implications for national security and homeland security~\cite{muntaka2025optimizing, rid2015attributing}.
Whereas its protocol justifiably enforces net neutrality and free speech~\cite{zantout2011i2p}, its anonymous design provides the technical capabilities required for conducting covert command-and-control (C2) activities, persistent orchestration channels, and communications that defy attribution efforts to a degree that makes it difficult to implement cyber attribution, cyber deterrence, and cyber warfare strategy~\cite{chen2025attribution, 
lin2012deterrence}. For practitioners and defenders operating in these
domains, understanding complex anonymous systems like
I2P is not optional. The network's architecture
determines what attribution techniques can and cannot
observe, and that boundary directly shapes the
feasibility of any response.

The predominant methodology for studying I2P has been
top-down empirical mapping for two
decades~\cite{egger2013i2p, biryukov2013trawling,
hoang2018empirical}: query the distributed network
database (NetDB), collect RouterInfo (RI) records from
floodfill nodes, enumerate observable peers, and
analyse LeaseSet distributions to characterise network
topology and behaviour. This approach has produced
genuine insight into the observable network. Its
structural limitation has gone unaddressed. The NetDB
is populated entirely by voluntary
publication~\cite{i2p_spec, muntaka2025resilience}.
A node that withholds its RouterInfo contributes no
record for any probe to retrieve. Increasing probe
frequency, expanding floodfill coverage, or refining
query scope all operate on the same directory that was
never populated. This is not a calibration problem.
It is an architectural property of the I2P protocol, and
it defines a space that every empirical mapping
technique in the literature is structurally incapable
of reaching.

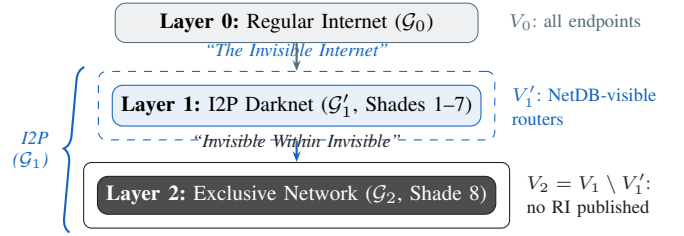
\begin{figure}[t]
\centering
\begin{tikzpicture}[
  every node/.style={font=\small},
  box/.style={draw=#1, fill=#1!10, rounded corners=4pt,
              minimum width=4.4cm, minimum height=0.50cm,
              text centered, text=black},
  arr/.style={-{Stealth[length=4pt]}, thick, #1}
]
\node[box=l0col, minimum width=4.8cm, minimum height=0.52cm]
     (l0) at (0,0)
     {\footnotesize\textbf{Layer 0:} Regular Internet
      ($\mathcal{G}_0$)};
\node[box=l1col, minimum width=4.2cm, below=0.55cm of l0]
     (l1)
     {\footnotesize\textbf{Layer 1:} I2P Darknet
      ($\mathcal{G}_1'$, Shades 1--7)};
\node[draw=l1col, dashed, rounded corners=3pt,
      inner sep=5pt, fit=(l1)]
     (l1box) {};
\node[above=2pt of l1box, font=\scriptsize\itshape, text=l1col]
     {``The Invisible Internet''};
\node[box=l2col, fill=l2col!80, text=white,
      minimum width=3.2cm, below=0.65cm of l1]
     (l2)
     {\footnotesize\textbf{Layer 2:} Exclusive Network
      ($\mathcal{G}_2$, Shade~8)};
\node[draw=l2col, rounded corners=3pt,
      inner sep=5pt, fit=(l2)]
     (l2box) {};
\node[above=2pt of l2box, font=\scriptsize\itshape, text=l2col]
     {``Invisible Within Invisible''};
\draw[arr=l0col] (l0.south)    -- (l1box.north);
\draw[arr=l1col] (l1box.south) -- (l2box.north);
\node[right=0.28cm of l0, font=\scriptsize,
      text=l0col, align=left]
     {$V_0$: all endpoints};
\node[right=0.28cm of l1, font=\scriptsize,
      text=l1col, align=left]
     {$V_1'$: NetDB-visible\\[1pt]routers};
\node[right=0.28cm of l2, font=\scriptsize,
      text=l2col, align=left]
     {$V_2 = V_1 \setminus V_1'$:\\[1pt]no RI published};
\draw[decorate,
      decoration={brace, amplitude=5pt, mirror},
      thick, l1col]
     ($(l1box.north west)+(-0.18cm, 0.05cm)$)
     --
     ($(l2box.south west)+(-0.18cm,-0.05cm)$)
     node[midway, left=7pt, font=\scriptsize\itshape,
          text=l1col, align=right]
     {I2P\\($\mathcal{G}_1$)};
\end{tikzpicture}
\caption{Three-layer network hierarchy. Layer~1
($\mathcal{G}_1'$) is the observable I2P darknet
(Shades~1--7). Layer~2 ($\mathcal{G}_2$) is the
Exclusive Network: routers that publish no RouterInfo,
invisible even within I2P.}
\label{fig:layers}
\end{figure}

In I2P, there is one such layer that plays this precise function. We label this layer the Exclusive Network and represent it graphically as Layer~2 in Fig.~\ref{fig:layers}. Nodes at this layer build tunnels, operate functional eepsites, and route garlic-encrypted traffic through the wider I2P infrastructure, hence utilizing bandwidth and tunnel capacity resources provided by others. These nodes are free riders: present in the true network yet wholly absent from any observable directory. The barrier to entry is low, allowing threat actors ranging from criminal groups to nation-state operators to deploy persistent covert infrastructure with no artifact exposed to any forensic investigator.

This gap carries practical consequences. For example, documented use cases such as the I2P-based remote access trojan (RAT), including I2PRAT (RATatouille)~\cite{sekoia2025i2prat}, leverage this behavior to facilitate persistent C2. Operational Relay Box (ORB) networks for nation-state campaigns~\cite{mandiant2024orb} 
also achieve this through jurisdictional dispersion of compromised relay infrastructure. Although two decades of I2P measurement efforts 
have not characterised this layer, Sections~\ref{sec:taxonomy} through~\ref{sec:results} seek to remedy this issue.

The incomplete nature of top-down empirical mapping sets an upper limit on how much can be learned from measurement alone. This work provides empirical evidence of that limitation, as well as an explanation for its underlying cause, thus providing motivation for the development of analytical techniques to complement the existing empirical methods. The contribution of this study is as follows:

\begin{enumerate}
\item We characterize the \emph{Exclusive Network}: I2P routers that host reachable services while publishing no RouterInfo, placing the host outside every RI-based mapping technique. We formalize it as the residual set $V_2 = V_1 \setminus V_1'$, situate it as the endpoint of an eight-class visibility taxonomy, and link it to documented C2 (I2PRAT) and nation-state ORB infrastructure.
\end{enumerate}

\section{Network Architecture and Formal Model}
\label{sec:background}

\subsection{Three-Layer Hierarchy}

The I2P (garlic network) is not simply a layer added to the existing Internet to provide anonymity \cite{i2p_spec}. Rather, the design of the I2P network creates a hierarchical nesting structure across three levels. We model this hierarchy using nested graphs, and the invisibility property arises directly as a result of the model architecture.

\begin{definition}[I2P Network Hierarchy]

Let $\mathcal{G}_0 = (V_0, E_0)$ be the Internet,
where $V_0$ is the set of reachable endpoints.
The I2P overlay will form a proper subgraph represented as
$\mathcal{G}_1 = (V_1, E_1)$, where
$V_1 \subset V_0$ represents all active I2P router
endpoints (Layer~1). Any RouterInfo-based mapping
of this overlay recovers only the observable
subgraph $\mathcal{G}_1' = (V_1', E_1')$, where
$V_1' \subseteq V_1$ comprises those routers that have published signed RouterInfo (RI) records to the
global NetDB managed by floodfill routers.
The Exclusive Network is the residual set $\mathcal{G}_2 = (V_2, E_2)$, where
$V_2 = V_1 \setminus V_1'$ (Layer~2), comprising
routers structurally absent from the NetDB.
\end{definition}

The non-emptiness of $V_2$ is an immediate result
of protocol design, not a measurement artefact.
I2P imposes no requirement on routers to publish
their RouterInfo (RI); directory participation is fully
voluntary~\cite{i2p_spec}. Thus, a router may join
$V_1$, build outbound tunnels as a client, and
contribute zero vertices to $\mathcal{G}_1'$.

\begin{proposition}[Structural Incompleteness]
\label{prop:incomplete}
For any RI-based mapping approach
$\mathcal{M}: \mathcal{G}_1 \to \mathcal{G}_1'$,
we have $\mathcal{G}_1' \subsetneq \mathcal{G}_1$
when at least one router operates in exclusive mode.
\end{proposition}

This follows from voluntary RI publication. From
a network science perspective, analogous structures
appear as the ``dark matter'' of complex
networks~\cite{barabasi1999}: entities that shape
system behaviour while absent from every observable
adjacency structure. The observable completeness
ratio $\rho = |V_1'| / |V_1|$ and its complement
$\xi = 1 - \rho$ quantify this gap, and are
evaluated empirically in Section~\ref{sec:results}.

\subsection{Garlic Routing and the NetDB}

I2P routes traffic through independent unidirectional
tunnel chains of depth $\ell$, where each relay knows only its immediate neighbours \cite{muntaka2026systemic}. Garlic
encryption bundles multiple payloads into a single
transmission, obscuring message boundaries from any
intermediate relay~\cite{i2p_spec}. The NetDB is a Kademlia-derived
distributed hash table (DHT)~\cite{maymounkov2002} whose storage
responsibility rotates daily through an XOR routing key expressed as: 
\begin{equation}
  \mathbf{rk}(d) =
    \mathrm{SHA\text{-}256}\!\left(H_d \oplus
    \mathrm{SHA\text{-}256}(\texttt{"yyyyMMdd"})\right)
  \label{eq:rk}
\end{equation}
where $H_d = \mathrm{SHA\text{-}256}(\mathbf{d}_\text{bytes})$
and $\mathbf{d}_\text{bytes}$ is the destination's
public-key. The floodfill whose 256-bit hash sits
XOR-nearest to $\mathbf{rk}(d)$ becomes the
designated storage node:
\begin{equation}
  f^*(d) = \mathop{\arg\min}_{f \in \mathcal{F}}
    \left[H_f \oplus \mathbf{rk}(d)\right]_\mathbb{Z}
  \label{eq:resp}
\end{equation}
where $[\cdot]_\mathbb{Z}$ denotes unsigned integer
interpretation. Equation~(\ref{eq:resp}) underpins
Method~C in Section~\ref{sec:methods}.

\begin{figure}[t]
\centering
\colorlet{s8bmid}{s8b!70}
\begin{tikzpicture}[
  node distance=0.26cm and 0.46cm,
  every node/.style={font=\scriptsize},
  rtr/.style={draw, circle, minimum size=0.52cm,
              fill=#1!14, draw=#1, thick},
  lbl/.style={font=\scriptsize\itshape, text=black!65},
  arr/.style={-{Stealth[length=3.5pt]}, thick, #1},
  dbox/.style={draw=l2col, fill=l2col!10,
               rounded corners=2pt,
               minimum width=1.55cm,
               minimum height=0.36cm, text centered}
]
\node[rtr=pkblue]           (cli) {\textbf{C}};
\node[lbl,below=0.05cm of cli] {Client};
\node[rtr=l1col, right=0.46cm of cli]  (o1) {$R_1$};
\node[rtr=l1col, right=0.36cm of o1]   (o2) {$R_2$};
\node[rtr=l1col, right=0.36cm of o2]   (o3) {$R_3$};
\node[lbl,below=0.04cm of o2] {Outbound ($\ell$-hop)};
\draw[arr=pkblue](cli)--(o1);
\draw[arr=l1col](o1)--(o2);
\draw[arr=l1col](o2)--(o3);
\node[rtr=s8bmid, below=0.78cm of o1]  (g)  {G};
\node[rtr=s8bmid, right=0.36cm of g]   (h1) {$H_1$};
\node[rtr=s8bmid, right=0.36cm of h1]  (h2) {$H_2$};
\node[rtr=l2col, right=0.36cm of h2,
      fill=l2col!80, text=white]         (ep) {\textbf{E}};
\node[lbl,below=0.04cm of h1] {Inbound ($\ell$-hop)};
\node[lbl,below=0.04cm of ep]
     {\textcolor{l2col}{\textbf{Shade~8 host}}};
\draw[arr=s8bmid](g)--(h1);
\draw[arr=s8bmid](h1)--(h2);
\draw[arr=l2col](h2)--(ep);
\draw[arr=c2r, dashed]
     (o3) to[out=-65,in=115] (g);
\node[font=\scriptsize\itshape, c2r,
      above right=-0.1cm and 0.02cm of o3] {Rdv};
\node[dbox, above=0.18cm of g, xshift=0.2cm]
     (ls) {LS: G-hash only};
\draw[-{Stealth[length=3pt]}, dashed, l1col](ls)--(g);
\node[c2r, font=\scriptsize, right=0.04cm of ep,
      align=left] {\textbf{Never}\\published};
\end{tikzpicture}
\caption{I2P inbound tunnel. The LeaseSet (LS)
publishes only the gateway hash \textbf{G}; the
Shade~8 hosting endpoint \textbf{E} is absent from
the LeaseSet and from the NetDB.
Rdv~=~rendezvous point; $\ell$~=~tunnel length.}
\label{fig:arch}
\end{figure}
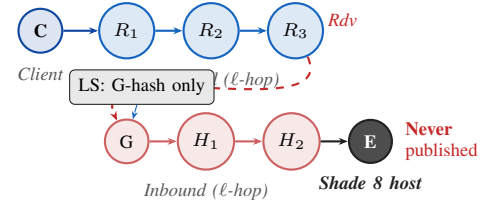

An eepsite's inbound LeaseSet publishes only the gateway hash and a tunnel identifier (Fig.~\ref{fig:arch}), disclosing nothing about the identity of the hosting router at the tunnel endpoint. An operator in the Exclusive Network withholds
the RouterInfo entirely, removing that endpoint
from $V_1'$ and placing it in $V_2$.

\section{Shade Taxonomy}
\label{sec:taxonomy}

The Shade Taxonomy formalises the visibility
gradient across I2P routers, making the boundary
between the observable network and the Exclusive
Network precise. Prior literature treats I2P as a binary distinction
between floodfill and non-floodfill routers for mapping purposes~\cite{egger2013i2p}. That framing is
insufficient. The taxonomy captures eight discrete
visibility classes derived entirely from observable
RouterInfo fields, demonstrating that structural
invisibility is the endpoint of a spectrum rather
than a binary property. Shades~1 through~7 reside
in Layer~1 ($V_1'$); Shade~8 defines Layer~2
($V_2$).

\begin{definition}[Shade Classifier]
Let $\kappa(r)$ denote the capabilities string of
router $r$, $\alpha(r) \in \{0,1\}$ indicate
whether a direct transport address is published,
$\iota(r) \in \{0,1\}$ indicate introducer
presence, and $\delta(r) \in \{0,1\}$ indicate
whether a RouterInfo record for $r$ is present in
the global NetDB. The shade class of $r$ is:
\begin{equation}
  \sigma(r) = \begin{cases}
    8 & \text{if } \delta(r) = 0\\
    f_\text{cap}\!\bigl(\kappa(r), \alpha(r),
               \iota(r)\bigr) & \text{otherwise}
  \end{cases}
  \label{eq:shade}
\end{equation}
where $f_\text{cap}$ maps observable capability
properties to shades 1 through 7 per
Table~\ref{tab:shades}.
\end{definition}

Shades~1 through~7 form a concealment progression
from Shade~1 (Beacon), which carries the floodfill
flag and anchors the NetDB, through firewalled,
introducer-only, and hidden-flag variants, to
Shade~7 (Phantom), which has no address and no
introducer yet still exists in the NetDB. All seven
satisfy $\delta(r) = 1$: present in the directory,
however hard to contact.

Shade~8 is categorically different. It satisfies
$\delta(r) = 0$: no NetDB record exists, and none
can be retrieved by any RouterInfo-based method
regardless of how many floodfills are probed. A Shade~7 router is hard to reach; a Shade~8 router is structurally absent from the NetDB. Hidden mode is the documented mechanism that suppresses RI publication~\cite{hoang2018empirical}; what distinguishes Shade~8 is its use as an operational service host: the published LeaseSet exposes only a tunnel gateway, never the hosting router (Fig.~\ref{fig:arch}). The classifier in Eq.~(\ref{eq:shade}) encodes this boundary by checking $\delta(r)$ before any capability inspection, since capability fields are
unavailable when no RouterInfo record exists.
The empirical validation of this property is the
subject of Section~\ref{sec:results}.

\begin{table}[t]
\centering
\scriptsize
\caption{Shade Taxonomy. $\kappa_f$: floodfill flag;
$\kappa_H$: hidden flag; $\kappa_U$: firewalled flag.
The Layer column refers to Fig.~\ref{fig:layers}.}
\label{tab:shades}
\setlength{\tabcolsep}{3pt}
\renewcommand{\arraystretch}{1.12}
\begin{tabular}{@{}clllc@{}}
\toprule
\textbf{Shade} & \textbf{Name} & \textbf{Criteria}
  & \textbf{C2 Role} & \textbf{Layer}\\
\midrule
1 & Beacon   & $\kappa_f$, $\alpha{=}1$
  & NetDB anchor   & 1\\
2 & Relay    & High-cap, $\alpha{=}1$
  & Traffic relay  & 1\\
3 & Passive  & Low-cap, $\alpha{=}1$
  & BW donor       & 1\\
4 & Cloaked  & $\kappa_U$, $\alpha{=}1$
  & Hidden relay   & 1\\
5 & Veiled   & $\alpha{=}0$, $\iota{=}1$
  & Covert relay   & 1\\
6 & Declared & $\kappa_H$, $\alpha{=}0$
  & Semi-hidden    & 1\\
7 & Phantom  & $\alpha{=}0$, $\iota{=}0$, $\delta{=}1$
  & Ghost node     & 1\\
\textbf{8} & \textbf{Exclusive}
  & $\boldsymbol{\delta{=}0}$
  & \textbf{Stealth C2}
  & \textbf{2}\\
\bottomrule
\end{tabular}
\end{table}


\section{Threat Model}
\label{sec:threat}

\subsection{Operational C2 Architecture}

A threat actor deploying a Shade~8 node as C2
infrastructure configures the router's
\texttt{router.config} file with parameters that
suppress all directory participation. Our exclusive
network script implements two progressively deeper
profiles~\cite{muntaka2026fiftyshades}:

\begin{lstlisting}[caption={Exclusive profile (10
parameters). Ghost profile adds 8+ more,\protect\\
including firewalled declaration and laptop-mode
identity rotation.}, label={lst:config},
breaklines=true, columns=fullflexible, captionpos=b]
router.isHidden=true
router.hiddenMode=true
i2np.udp.addressSources=      # empty
i2np.ntcp2.autoip=false
router.floodfillParticipant=false
router.maxParticipatingTunnels=0
router.sharePercentage=0
router.enablePeerTest=false
router.dynamicKeys=true        # ephemeral identity
i2np.udp.requireIntroductions=true
\end{lstlisting}

With these settings, the router publishes no RI to
any floodfill, refuses to relay traffic for other
nodes, disables peer-testing probes that would reveal
reachability, and rotates its cryptographic identity
on every restart without affecting eepsite keys,
which are stored separately in \texttt{eepPriv.dat}.
The hosted eepsite remains reachable to partners who
hold the b32 address.

\subsection{Connection to Documented Malware}

Remote access trojan variants that exploit the I2P architecture, for example I2PRAT (RATatouille)~\cite{sekoia2025i2prat} implement this paradigm explicitly. The backdoor initiates a connection to the local I2P proxy using the SAM bridge (port 7656) and engages in communication with the C2 eepsite having its b32 address hardcoded into the malicious executable code. In cases where the server-side node is set up as the Exclusive Network (Shade 8), the outgoing network traffic from the victimized machine will look like regular I2P traffic \cite{sekoia2025i2prat} ~\cite{kaspersky2021mata}.
Attribution then requires either binary forensics
on the implant or comprehensive traffic correlation
across the entire I2P overlay.

\subsection{Operational Relay Box Parallel}

Operational Relay Box (ORB) networks, attributed to
multiple Chinese APT groups, construct multi-hop
proxy chains through compromised small office home office (SOHO) routers and cloud infrastructure to obscure operation origins~\cite{mandiant2024orb}. Where ORB achieves
unattributability through jurisdictional dispersion
of relay infrastructure, the Exclusive Network achieves it through protocol-level directory non-publication. Both instantiate
$G_\text{dark} \subset G$: an operational subgraph
that contributes to network behaviour while remaining
absent from every observable directory. This shared structure positions the model developed in Section~\ref{sec:background} as
directly applicable to the structural analysis of
ORB-class architectures.


\section{Empirical Methodology}
\label{sec:methods}

\subsection{Three-Node Testbed}

The testbed comprised three nodes running
Ubuntu~24.04~LTS with I2P~2.12.0 (API~0.9.69).
All configuration scripts are available on
GitHub~\cite{muntaka2026fiftyshades}.

\smallskip
\noindent\textbf{VM1 (Exclusive Host).} Router
$\mathcal{H}_1 = \texttt{PB5dY5gvdEpj\ldots}$,
configured via \texttt{exclusiveStealth-network.sh}
(ghost profile, Listing~\ref{lst:config}). Hosts
eepsite \texttt{sid001.i2p} with custom TLD routing
managed by \texttt{customtld-manager.sh}.

\noindent\textbf{VM2 (Authorised Partner).} Router
$\mathcal{H}_2 = \texttt{6FRyiaaN\ldots}$,
configured via \texttt{setup-i2p-proxy.sh} with
SOCKS5 access to VM1's eepsite. VM2 holds VM1's
b32 address obtained through an out-of-band channel.

\noindent\textbf{VM3 (Adversary Scanner).} Runs
\texttt{node-lookup.py} with knowledge of
$\mathcal{H}_1$'s router hash but no knowledge of
its b32 eepsite address, and separately runs
\texttt{b32-lookup.py} with knowledge of the b32
address but no knowledge of the hosting router hash.
These two complementary probes simulate an adversary
approaching from either direction, together
constituting the empirical test of Shade~8
structural invisibility.

\subsection{NetDB Measurement Approach}

RouterInfo records are collected from the local
NetDB directory
(\texttt{\textasciitilde/.i2p/netDb/}) by parsing
binary \texttt{routerInfo-*.dat} files via the
POSIX \texttt{strings} utility, extracting
capabilities, transport addresses, version strings,
and known-peer counts. The I2P console API at
\texttt{127.0.0.1:7657/netdb} provides a second
source. Our snapshot consisted of 3,272~RI entries, with 1,536 (46.9\%) having the floodfill capability flag set, consistent with established
I2P measurement practice~\cite{egger2013i2p,
muntaka2025mapping}.

\subsection{Attribution Methods}
\label{sub:methods}

The attribution framework consists of five
approaches: three incorporated into
Algorithm~\ref{alg:shade} (local NetDB inspection,
console cache query, and floodfill probe expansion)
and two described below. Method~A is a control
verification from the hosting VM; Methods~B+D
and~C are the adversarial approaches applied
by VM3.

\noindent\textbf{Method~A (b32 Derivation).}
On the hosting VM, the canonical b32 address follows
directly from \texttt{eepPriv.dat}:
\begin{equation}
  b32(r) = \mathrm{Base32}\!\bigl(
    \mathrm{SHA\text{-}256}(
      \mathbf{d}_r[0:d_s])\bigr)
  \label{eq:b32}
\end{equation}
$d_s = 387 + L$, where $L = \text{uint16\_be}(
\mathbf{d}_r[385:387])$ is the key-certificate
length. Confirmed on I2P~2.12.0: cert type~5,
$d_s = 391$ bytes,
$b32 = \texttt{poitcahygw7f2zz7\ldots}$.

\noindent\textbf{Method~B+D (Gateway Scan).}
All known active LeaseSets are inspected for Lease
entries whose gateway hash prefix matches the
target. With $\ell$-hop tunnels ($\ell \geq 1$),
a gateway match indicates routing participation,
not hosting.

\noindent\textbf{Method~C (XOR Routing Key).}
Using Eqs.~(\ref{eq:rk}) and (\ref{eq:resp}), for
each known b32 address, the responsible floodfill
is identified. If the target matches, it stores
that LeaseSet per protocol.

Algorithm~\ref{alg:shade} formalises the Shade
classification; Algorithm~\ref{alg:xor} formalises
Method~C.

\begin{algorithm}[t]
\caption{Shade Classification Protocol}
\label{alg:shade}
\begin{algorithmic}[1]
\Input Router hash $h$, local NetDB $\mathcal{D}$,
       floodfill set $\mathcal{F}$, batch size $b$
\Output Shade class $\sigma(h) \in \{1,\ldots,8\}$
\If{$h \in \mathcal{D}$}
  \State Extract $(\kappa, \alpha, \iota)$ from
         local RI
  \State \Return $f_\text{cap}(\kappa, \alpha, \iota)$
         \Comment{Shades 1--7}
\EndIf
\State Query console: $\text{RI} \gets
       \texttt{/netdb?r=}h$
\If{$\text{RI} \neq \emptyset$}
  \State \Return $f_\text{cap}(\kappa, \alpha, \iota)$
         from RI
\EndIf
\For{$i \gets 0$ to $|\mathcal{F}|$ step $b$}
  \For{$f \in \mathcal{F}[i:i{+}b]$}
    \State Probe: \texttt{/netdb?r=}$f$
           \Comment{Expand console view}
  \EndFor
  \State $\text{RI} \gets \texttt{/netdb?r=}h$
  \If{$\text{RI} \neq \emptyset$}
    \State \Return $f_\text{cap}(\kappa, \alpha,
                   \iota)$ from RI
  \EndIf
\EndFor
\State \Return $8$
       \Comment{Shade~8: $\delta(h)=0$ confirmed}
\end{algorithmic}
\end{algorithm}

\begin{algorithm}[t]
\caption{XOR Routing Key Association (Method~C)}
\label{alg:xor}
\begin{algorithmic}[1]
\Input Target hash $H_t$, eepsite set
       $\mathcal{S}$, floodfill map $\mathcal{F}$
       (hash $\to$ bytes)
\Output $\mathcal{R} \subseteq \mathcal{S}$:
        eepsites for which $H_t$ is responsible
\State $m_k \gets \mathrm{SHA256}(
       \text{UTC\_date\_as\_yyyyMMdd})$
\State $\mathcal{R} \gets \emptyset$
\For{$s \in \mathcal{S}$ with known b32}
  \State Decode $H_s$ from b32 prefix
  \State $\mathbf{rk}_s \gets
         \mathrm{SHA256}(H_s \oplus m_k)$
  \State $d_t \gets [H_t \oplus \mathbf{rk}_s
         ]_\mathbb{Z}$
  \State $\texttt{closest} \gets \textbf{true}$
  \For{$f \in \mathcal{F},\; f \neq H_t$}
    \If{$[H_f \oplus \mathbf{rk}_s]_\mathbb{Z}
        < d_t$}
      \State $\texttt{closest} \gets \textbf{false}$;
             \textbf{break}
    \EndIf
  \EndFor
  \If{\texttt{closest}}
    \State $\mathcal{R} \gets \mathcal{R}
           \cup \{s\}$
  \EndIf
\EndFor
\State \Return $\mathcal{R}$
\end{algorithmic}
\end{algorithm}


\begin{table}[t]
\centering\footnotesize
\caption{Per-router classification outcomes. ``NetDB RI'' is whether a RouterInfo was retrieved across the local, console, and 500-probe sources; ``absent (0/500)'' denotes zero hits at every checkpoint. Full console output in Figs.~\ref{fig:nodelookup},~\ref{fig:shade1beacon}, and~\ref{fig:shade2relay}.}
\label{tab:results}
\setlength{\tabcolsep}{4pt}
\renewcommand{\arraystretch}{1.2}
\begin{tabular}{@{}llccll@{}}
\toprule
\textbf{Router} & \textbf{Caps} & \textbf{Kn.\ Rtrs} & \textbf{LeaseSets} & \textbf{NetDB RI} & \textbf{Shade}\\
\midrule
$\mathcal{H}_1$ & ---          & ---     & --- & absent (0/500) & \textbf{8}\\
$\mathcal{F}_1$ & \texttt{XfR} & 7{,}778 & 213 & present        & 1\\
$\mathcal{R}_1$ & \texttt{XR}  & 0       & 0   & present        & 2\\
\bottomrule
\end{tabular}
\end{table}

\section{Experimental Results}
\label{sec:results}

\subsection{Shade~8 Empirical Demonstration}

Algorithm~\ref{alg:shade} ran on VM3 targeting
$\mathcal{H}_1 = \texttt{PB5dY5gvdEpj\ldots}$,
with $b = 5$ and $|\mathcal{F}| = 1{,}536$:

\begin{enumerate}
\item \textbf{Local NetDB} (3,272 RI files): no hit.
\item \textbf{Console cache}: no hit.
\item \textbf{Floodfill probe} (500 floodfills,
  batches of 5, re-check after each): no hit at
  any checkpoint.
\end{enumerate}

The Shade~8 criterion is satisfied:
\begin{equation}
  \neg\,\mathrm{RI}_\text{local}(\mathcal{H}_1)
  \;\wedge\;
  \neg\,\mathrm{RI}_\text{console}(\mathcal{H}_1)
  \;\wedge\;
  \bigwedge_{f \in \mathcal{F}_{500}}
    \neg\,\mathrm{RI}_f(\mathcal{H}_1)
  \label{eq:shade8}
\end{equation}

\begin{figure}[!t]
\centering
\includegraphics[width=0.60\columnwidth]{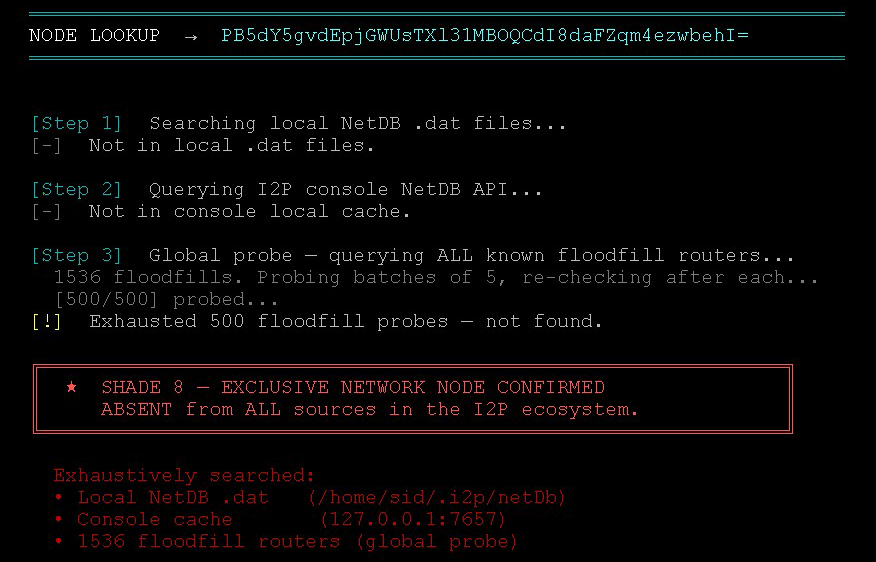}
\caption{Shade~8 classification output from
\texttt{node-lookup.py}. After 500 floodfill
probes from a pool of 1,536, router $\mathcal{H}_1$
(\texttt{PB5dY5\ldots}) produces zero NetDB hits,
confirming Layer~2 exclusive status.}
\label{fig:nodelookup}
\end{figure}

Fig.~\ref{fig:nodelookup} confirms zero NetDB hits across all three sources, with per-router outcomes summarized in Table~\ref{tab:results}. Simultaneously, VM2 maintained continuous access to \texttt{sid001.i2p}, confirming that structural
invisibility and operational functionality coexist
in Layer~2.

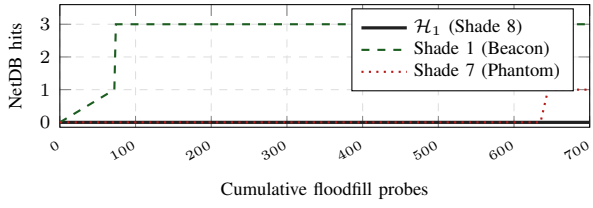
\begin{figure}[t]
\centering
\begin{tikzpicture}
\begin{axis}[
  width=0.97\columnwidth, height=3.2cm,
  xlabel={\scriptsize Cumulative floodfill probes},
  ylabel={\scriptsize NetDB hits},
  xmin=0, xmax=700,
  ymin=-0.15, ymax=3.6,
  ytick={0,1,2,3},
  xtick={0,100,200,300,400,500,600,700},
  xticklabel style={font=\tiny, rotate=30,
                    anchor=north east},
  yticklabel style={font=\scriptsize},
  grid=major, grid style={dashed, gray!25},
  legend pos=north east,
  legend style={font=\scriptsize,
                cells={anchor=west},
                row sep=-2pt},
  every axis plot/.append style={thick}
]
\addplot[l2col, very thick, solid]
  coordinates {(0,0)(100,0)(200,0)(300,0)(400,0)
               (500,0)(600,0)(700,0)};
\addlegendentry{$\mathcal{H}_1$ (Shade~8)};
\addplot[s1g, dashed, thick]
  coordinates {(0,0)(72,1)(73,2)(74,3)(700,3)};
\addlegendentry{Shade~1 (Beacon)};
\addplot[s8b, dotted, thick]
  coordinates {(0,0)(635,0)(645,1)(700,1)};
\addlegendentry{Shade~7 (Phantom)};
\end{axis}
\end{tikzpicture}
\caption{NetDB hit count vs.\ cumulative floodfill
probes for three router types. $\mathcal{H}_1$
(Shade~8, Layer~2) remains absent across all 500
probes; observable Layer~1 nodes are found within
tens to hundreds of probes.}
\label{fig:results}
\end{figure}

Fig.~\ref{fig:results} plots NetDB hit count
against cumulative probes for three router classes,
confirming $\mathcal{H}_1$ remains in
$V_2 = V_1 \setminus V_1'$ across all 500 probes.

\subsection{Method~C: Floodfill Association}

\begin{figure}[!t]
\centering
\includegraphics[width=0.60\columnwidth]{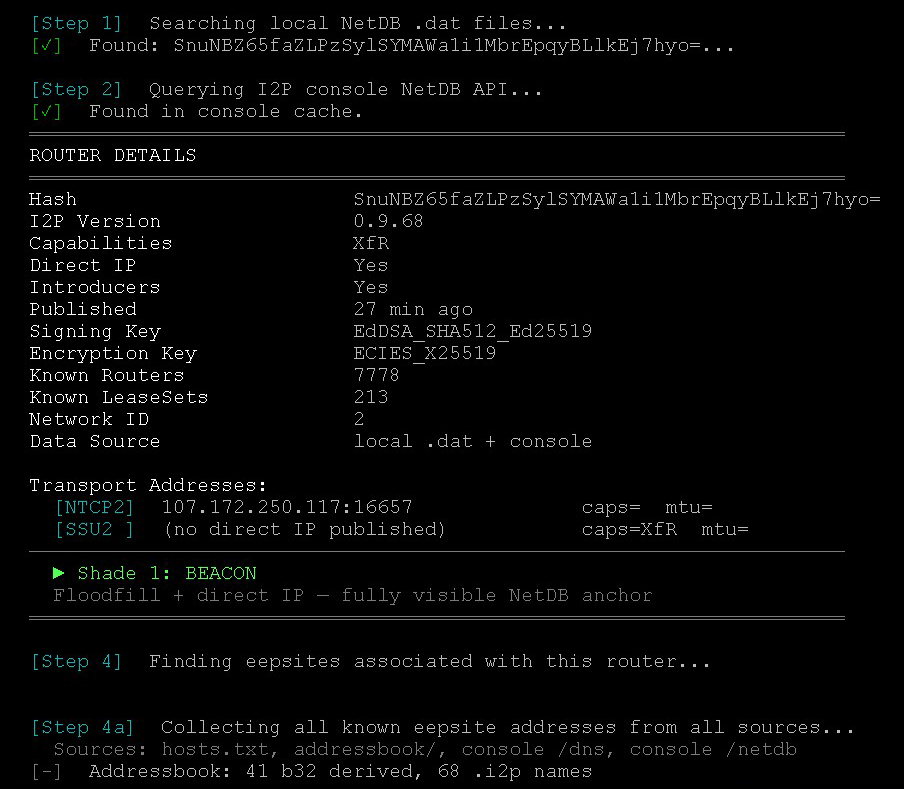}
\caption{Shade~1 (Beacon) classification output
for router $\mathcal{F}_1$ (\texttt{SnuNBZ65\ldots}).
Known Routers:~7,778; Known LeaseSets:~213;
caps \texttt{XfR}. All five detection methods
return positive results, confirming full
observability in Layer~1.}
\label{fig:shade1beacon}
\end{figure}

Algorithm~\ref{alg:xor} targeted floodfill
\texttt{SnuNBZ65faZL\ldots} (caps~\texttt{XfR},
I2P~0.9.68, \texttt{107.172.250.117:16657},
Known LeaseSets:~213, Fig.~\ref{fig:shade1beacon})
against 1,536 floodfills and 172 candidate b32
addresses. It identified
\texttt{2yxn3eio\ldots.b32.i2p} as the one address
in the candidate set for which this floodfill is the XOR-nearest known storage node.

\begin{figure*}[!t]
\centering
\includegraphics[width=0.58\textwidth]{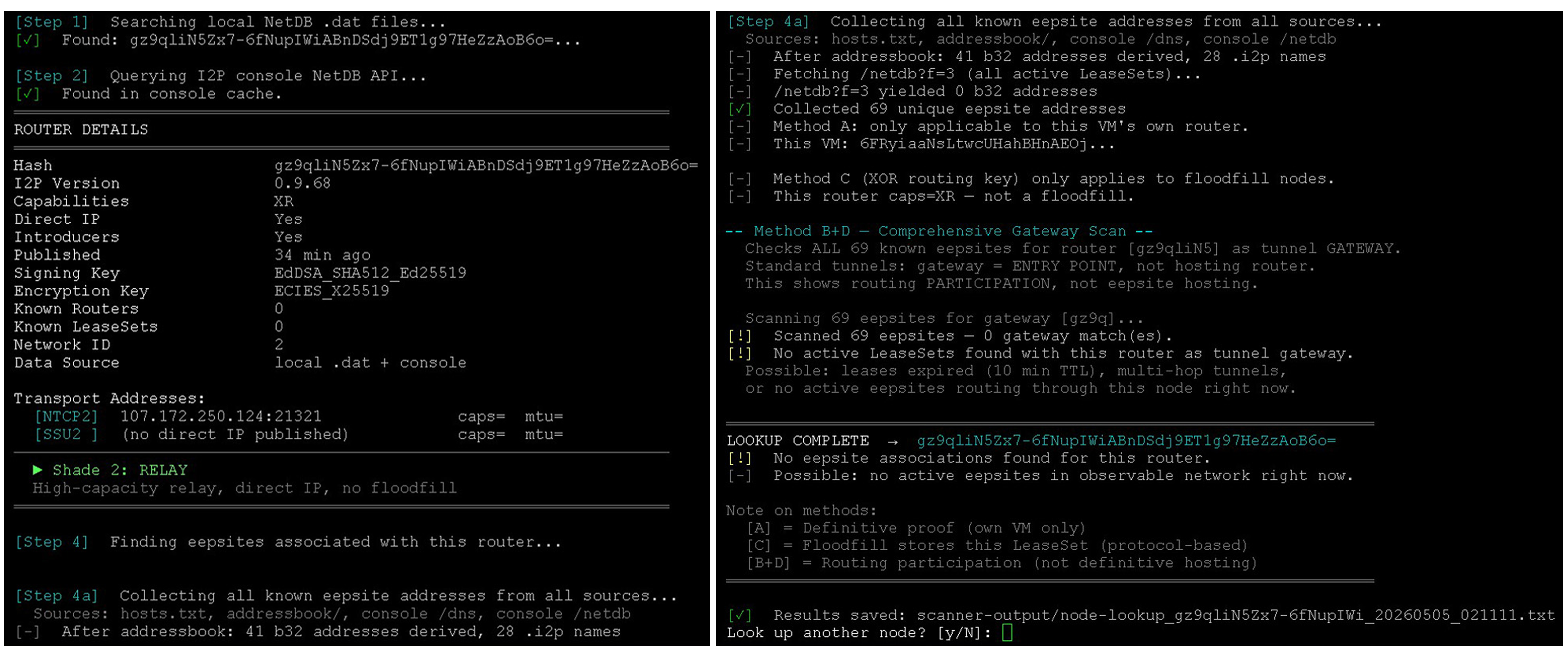}
\caption{Shade~2 (Relay) classification for
router $\mathcal{R}_1$
(\texttt{gz9qliN5\ldots}, caps \texttt{XR}).
Left: Known Routers:~0, Known LeaseSets:~0,
direct IP published, Shade~2: Relay. Right:
Methods~B and~D scan 69 known eepsites and return zero gateway associations. Despite full
tunnel routing participation, $\mathcal{R}_1$
carries no service descriptors and occupies
Layer~1 Shade~2.}
\label{fig:shade2relay}
\end{figure*}

For comparison, Fig.~\ref{fig:shade2relay} shows
relay node \texttt{gz9qliN5Zx7\ldots}
(caps~\texttt{XR}, Shade~2). Zero known routers
and zero LeaseSets confirm no NetDB storage
participation; Methods~B and~D scan 69 known eepsites and return zero gateway matches.
Despite full tunnel routing participation, this
node carries no service descriptors, placing it
firmly in Layer~1 Shade~2 as a structural
contrast to $\mathcal{H}_1$.

\section{Discussion}
\label{sec:discussion}

\subsection{Attribution and the Observable Graph Limit}

The result of Eq.~(\ref{eq:shade8}) demonstrates
a hard epistemic boundary for NetDB-based
attribution. Every technical methodology applied
to $\mathcal{G}_1'$, including IP geolocation,
RouterInfo fingerprinting, and floodfill
enumeration, is bounded within $V_1'$. A Shade~8
operator is necessarily in $V_2 = V_1 \setminus
V_1'$, beyond this boundary by construction.
Identifying actors in $V_2$ requires either
endpoint forensics or global traffic correlation
across all relay hops in $\mathcal{G}_1$, neither
tractable at operational tempo for most defenders.

This mirrors the structural challenge of ORB
attribution~\cite{mandiant2024orb}: in both cases,
the observable network is a proper subgraph of the
true operational topology. I2P achieves invisibility
through protocol-level non-publication; ORB achieves
it through jurisdictional dispersion of compromised
relay infrastructure.

\subsection{Implications for Cyber Deterrence}

Deterrence theory requires that a credible
retaliatory or sanctioning response be made possible
by attribution~\cite{lin2012deterrence}. Layer~2
I2P infrastructure severs this chain at its
foundation, leaving a C2 operator using an exclusive
node with no credible technical-attribution-based
deterrence. Deterrence strategies must therefore
shift from infrastructure-based attribution to
behavioural attribution: the analysis of targeting
patterns, exploit tooling, and operational rhythms,
which is less dependent on the visibility of
individual network nodes.

\subsection{Network Science: Dark Vertices}

The Shade Taxonomy contributes a vertex-visibility
classification to the network science of anonymous
systems. Shade~1 routers exhibit high betweenness
centrality in $\mathcal{G}_1'$; they are the primary
carriers of routing state. Shade~8 routers have zero
degree in $\mathcal{G}_1'$ despite positive degree
in the true graph $\mathcal{G}_1$, analogous to
``dark nodes'' in social network analysis where
influential actors are absent from observable contact
graphs~\cite{barabasi1999}. The complement
$\xi = 1 - \rho$ bounds the fraction of the true
topology that remains structurally inaccessible
regardless of measurement methodology.

\subsection{Dual-Use Dimension}

The civil applications of I2P, which safeguard journalists, dissidents, and whistle-blowers in
repressive or censored environments, are firmly
established and independent of the threat surface analysed here. The dual-use nature of the network cannot be addressed through technical restriction
of Layer~2 features without simultaneously undermining those protections. Policy should emphasise behavioural attribution capability rather than protocol interdiction.

\section{Conclusion}
\label{sec:conclusion}

In I2P, a sublayer exists where nodes consume routing resources without any directory record. Minimal configuration renders a standard I2P router invisible to every observable directory while remaining fully operational as a covert host. Two decades of empirical research have undercharacterized this layer: the directory is never populated, and no increase in probe coverage can retrieve what is never stored.

The empirical result is unambiguous: 500 sequential floodfill probes from a pool of 1,536, applied through five attribution methods, returned zero NetDB hits for $\mathcal{H}_1$ while its hosted eepsite remained continuously accessible. The bound $\xi = 1 - \rho$ is a hard protocol-design limit. Actors who understand this, such as in I2PRAT~\cite{sekoia2025i2prat}, use techniques to operate where
directory-based attribution is structurally impossible. The same property $G_\text{dark} \subset G$ defines ORB infrastructure~\cite{mandiant2024orb}; both lie beyond directory-based empirical methods.

The primary implication is methodological:
understanding the Exclusive Network demands formal
analytical approaches independent of directory
observation, and this study establishes precisely
why empirical mapping alone is insufficient.
Future research will expand the Shade Taxonomy
to i2pd and I2P+ variants~\cite{muntaka2026systemic},
investigate timing-analysis techniques for
Shade~8 de-anonymisation, and deploy I2PRAT
(RATatouille)~\cite{sekoia2025i2prat} as an
active C2 implant against an Exclusive Network
node to validate attribution-resistance under
adversarial conditions. The $G' \subset G$
incompleteness framework further offers a
structural foundation for graph-theoretic
modelling of ORB architectures, where nation-state
actors achieve comparable unattributability
through jurisdictional dispersion rather than
protocol-level non-publication.

\section*{Acknowledgment}
The authors acknowledge the support of Multi-domain
and Information Operations, Resilience and Anonymity
Groupe (MIRAGe-UC) research group at the University
of Cincinnati.

\bibliographystyle{IEEEtran}
\bibliography{ref}

\balance

\end{document}